# SPEC application for achieving inelastic X-ray scattering experiment in the SSRF

LAN Xu-Ying(兰旭颖)[1,2] YANG Ke(杨科)[1] LIANG Dong-Xu(梁东旭)[1] YAN Shuai(闫帅)[1] MAO Cheng-Wen(毛成文)[1] LI Ai-Guo(李爱国)[1] WANG Jie(王劼)[1]

[1] Shanghai institute of applied physics, Chinese academy of sciences, Shanghai 201204, China

[2] University of Chinese academy of sciences, Beijing 100049, China

**Abstract:** In order to carry out inelastic X-ray scattering (IXS) experiment at BL15U1 beamline of Shanghai Synchrotron Radiation Facility (SSRF), the data acquisition and control system based on SPEC software has been developed. The IXS experimental method needs linkage control of monochromator, silicon drift detector (SDD) and ionization chamber on continuous segment-scan mode with variable step size, and gains the data of energy, spectrum and light intensity synchronously. A method is presented for achieving this function which was not realized only by using SSCAN of Experimental Physics and Industrial Control System (EPICS). This paper shows work details including control system description, SPEC configurations for EPICS devices, macro definitions and applications in the BL15U1. An IXS experiment was executed by using the SPEC control system, its results prove that the method is feasible to perform the experiment.

## 1 Introduction

The IXS is a powerful technique based on synchrotron radiation facility. IXS can provide important dynamic information of excitations of maeterials, such as phonon, plasmon, exiton, d-d trasition etc[1-3]. We successfully build an IXS system at BL15U1 of SSRF. In this paper, the linkage control system of monochromator, SDD and ionization chamber are established for IXS. Because the interested inelastic peak needs a fine scanning while the elastic peak can be in a rough scanning course, this control realizes monochromator arbitrarily regulates energy on continuous segment mode with variable step size, the SDD collects photon counts and ionization chamber measures the photon flux synchronously. However, the SSCAN of EPICS which we used to scan energy is not suitable for arbitrary continuous segment scan with variable step size, it does scan in fixed step size without continuous segment[4-6]. In addition, users tend to systematically control devices in one interface, but one has to control different devices with different interfaces by using original EPICS software system, it leads to inconvenient operations. These have decreased practicality of experiment methods.

We apply SPEC software technique to build a data acquisition and control system for satisfying above-mentioned requirements. This is mainly based on three features of SPEC: first, it can support for EPICS distributed real-time instrument control systems and devices using the standard EPICS record [7-8]; second, macro definitions allow any final user to easily modify and adjust the control system to any specific requirement[9-11]; third, all experimental operations can be done systematically in one interface.

In this work, the whole control system for the IXS experiment is developed in the BL15U1. Hardware system based on commercially available components is described. In software aspects, macro configurations for EPICS devices, macros for control and data acquisition are all given. An IXS experiment based on this control system is detailed.

## 2 Control and data acquisition system
### 2.1 System structure

The work was developed on the basis of existing hardware at BL15U1 in SSRF. Experimental instrument based on a distributed control system is integrated with motors, detectors and some other devices. They are assembled from commercially available components. Motors are produced by KOHZU. Detectors include OKEN S1194A1 ionization chamber which is used to measure the incident light intensity and a Vortex SDD to collect photon counts. The double-crystal monochromator is used for scanning energy.

The whole distributed control system includes operator interface (OPI) level, input/output controller (IOC) level, LAN connecting the OPI and IOC, controlled devices[12-14]. The OPI level is any type of PC with Linux/Windows. EPICS soft IOC, SPEC and some other extensions run on the PC. In terms of IOC level, the controller could be any acquisition/control device, such as PXI and VME platforms with the appropriate cards, or a PC. This system can set up an EPICS IOC, which controls the communication with the devices/drivers and performs any further data processing. Controlled devices are motors, ionization chamber, SDD as we mentioned above. The general diagram of the data acquisition and controller system is presented in Fig.1.

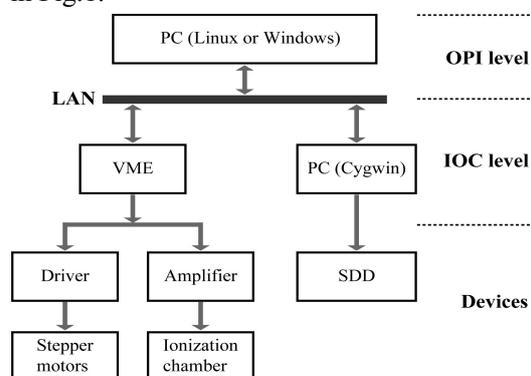

Fig.1 General diagram of data acquisition and control system

### 2.2 Macro configuration

In the control system which combined SPEC and EPICS, for building macro control, the macro hardware should be configured firstly.

We configure the motors which use the standard EPICS motor record. The motor configuration includes motor name, number of motor, PV of each motor and device type[15]. We use SPEC motor as 'EPICS_M2'. The PV prefix is set in the config file, such as X15U1:EH:, and appended MAN7:Z by the optional 'Generic parameter 1', spec creates PV from that prefix along the lines of: X15U1:EH:MAN7:Z.VAL. The device type is set as 'EPICS motor controller'.

Counter/timers using the standard EPICS scaler record is employed, some process variables are also configured as counters[16]. The scaler SIS3820 is selected as a master timer with counter channels supported by the standard EPICS scaler record, the device type of it is set as 'EPICS_SC'. Some process variables of SDD channels are also used as counters by configured as 'EPICS PV as Counter'. For the process variable counter, the device type is selected with 'EPICS_PV'.

### 2.3 Macro scan

For realizing energy scan on arbitrary continuous segment mode with variable step size, it needs standard macros and user defined macros, such as motor.mac, plot.mac, energy.mac and so on[17].

There are many different standard scans available. Absolute-position motor scans such as 'ascan'. 'a2scan'. Relative-position motor scans are 'dscan', 'd2scan'. We use the 'dscan' and the 'd2scan' in the experiment, because using the relative-position scans the motors return to their starting positions after the last point. 'dscan(d2scan)'/'ascan(a2scan)' are used not only for deciding sample position, but also for building user defined macro which can regulate energy. 'energy.mac' is used for an energy-selecting monochromator, such as 'escan' which is specifically for energy scan.

User defined macros are also built. 'gscan' allows users to scan one-motor through multiple regions with variable stepsizes for each one. For example, 'gscan motor 100 1 200 0.2 300 1 400 5 600  1', this means scan motor from 100 to 200 with a step of 1.0, followed by 0.2-stepping to 300, 1.0-stepping to 400, 5.0-stepping to 600, with an integration time of 1 sec for each points.

Scan macros which can realize linkage control for energy scan on continuous segment-scan mode are programmed using the above defined macros. One example is shown in Appendix.

Any specific function of IXS experiment can be easily built by programming user defined macros with C code.

## 3 Experiments and results

The IXS experiment was carried out in the BL15U1. In this experiment, energy is changed by adjusting the Bragg angle motor of monochromator, while the IXS spectrum is collected by the SDD synchronously. The K-B mirrors system is used to provide a focused beam with a spot size of 3×3μm. The sample is a diamond mounted on a 7-axis sample stage driven by stepper motors. First, we use 'dscan' and 'd2scan' to move sample stage motors for deciding sample position. Second, `setup_mca` is used to configure a new multi-channel analysis (MCA) and setup the region of interest (ROI) of SDD. In this step, the EPICS MCA PV and EPICS DXP PV should be known, the ROI can also be set by macro 'mca_rois'. We use `mca_on` and 'ct' to count photons of ROI. Third, 'escan' and 'gscan' is used to scan monochromator energy and collect SDD data synchronously, user defined macros are also used which realize linkage control and data acquisition on request. For the custom defined macros, epics_get(), epics_put() and epics_par() can be used for communicate with any PV in the EPICS. The result of this operation is a file containing all the data collected during a scan: energy, live-time, light intensity and photon count, which can be processed by a well-know software package PyMca. Such a file can also be used as input for further calculations. One of the experiment results which is Be inelastic peak is shown in Fig. 2.

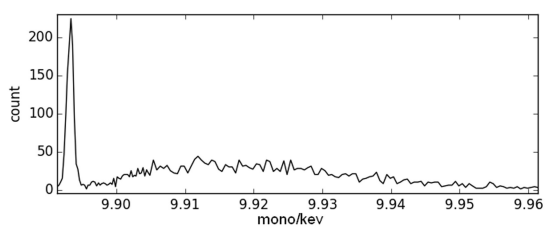

Fig.2 Inelastic peak of Be

## 4 Conclusions

This paper has presented SPEC software is used for IXS experiment in the EPICS environment at BL15U1 of SSRF. EPICS motors and scaler have been configured in the SPEC. Some macros are defined and their applications are given. Based on these works, we achieve linkage control of monochromator, SDD and ionization chamber on continuous segment-scan mode with variable step size, and data acquisition synchronously. An IXS experiment was carried out with this control and data acquisition system, the results prove this method's suitability and performance. Additionally, the technique provided in this paper can also be introduced to other control system.

**Acknowledgments**
This work was supported by the National Natural Science Foundation of China (NSFC) (11205237).

**Appendix:**

```
### scan with fixed step size
umv mono 10.194
for ( ; ; ) {
    for (idx=0; idx< 10; idx++) {
        newfile 1
        umv mono 10.194; dscan mono -0.025 0.015 28 30
    }
    umvr Z 0.1
    for (idx=0; idx< 25; idx++) {
        newfile 2
        umv mono 10.194; dscan mono -0.025 -0.011 28 30
    }
}

###scan with variable step size
umv mono 9.8934
for ( ; ; ) {
    for (idx=0; idx<2; idx++) {
        newfile 3
        umv mono 9.8934; gscan mono 9.8914 0.00025 9.9044 0.0005 9.9614 20
    }
    umvr Z 0.1
    for (idx=0; idx< 2; idx++) {
        newfile 4
      umv mono 9.8934; gscan mono 9.8914 0.00025 9.9044 0.0005 9.9614 20
    }
```